\documentclass[DIV=calc, paper=letter, fontsize=11pt]{scrartcl}	 

\usepackage[]{units}

\usepackage{lipsum} 
\usepackage[english]{babel} 
\usepackage[protrusion=true,expansion=true]{microtype} 
\usepackage{amsmath,amssymb,amsfonts,amsthm,bm,bbm} 

\usepackage[svgnames, table]{xcolor} 

\usepackage[hang, small,labelfont=bf,up,textfont=it,up]{caption} 
\usepackage{booktabs}
\usepackage{fix-cm}
\usepackage[]{units}

\usepackage{nicematrix} 

\usepackage{tabularx} 

\usepackage[colorlinks=true, linkcolor=blue, citecolor=blue, filecolor=blue,
runcolor=blue, urlcolor = black]{hyperref} 

\usepackage[capitalize, noabbrev]{cleveref}

\usepackage{fullpage}
\usepackage{sectsty} 

\usepackage{fancyhdr} 
\usepackage{lastpage}

\usepackage{nicefrac}

\usepackage{float}

\usepackage[utf8]{inputenc}
\usepackage{comment}

\usepackage[
backend=bibtex, 
style=numeric-comp,
sorting=none
]{biblatex}

\makeatletter
\newcommand{\multicolinterrupt}[1]{
\setcounter{tempcolnum}{\col@number}
\end{multicols}
#1
\begin{multicols}{\value{tempcolnum}}
}
\makeatother

\definecolor{issuePJA_color}{rgb}{1.0,0.0,0.0}

\definecolor{commentPJA_color}{rgb}{1.0,0.0,0.8}

\definecolor{rev_color}{rgb}{0.6,0.0,0.0}

\definecolor{atz_table1}{rgb}{0.85, 0.85, 0.85}
\definecolor{atz_table2}{rgb}{0.8, 0.8, 0.8}
\definecolor{atz_table3}{rgb}{0.75, 0.75, 0.75}

\definecolor{atzmlabel_color}{rgb}{0.0,0.0,0.0} 
\newcommand{\atzmfont}{\fontsize{8}{9}} 
\newcommand{\atzmlabel}[1]{\mbox{\atzmfont \textcolor{atzmlabel_color}{#1}}}

\newcommand{\mb}[1]{\mathbf{#1}}
\newcommand{\bsy}[1]{\boldsymbol{#1}}

\newcommand{\dvol}{\delta{V}} 

\usepackage{graphicx}
\DeclareGraphicsExtensions{.png}

\usepackage{lettrine} 
\newcommand{\initial}[1]{ 
\lettrine[lines=3,lhang=0.3,nindent=0em]{
\color{DarkGoldenrod}
{\textsf{#1}}}{}}

\usepackage{caption}

\DeclareOldFontCommand{\bf}{\normalfont\bfseries}{\mathbf}

\usepackage{graphicx}

\usepackage{tikz}
\usepackage{float}

\usepackage{titling} 

\newcommand{\HorRule}{\color{DarkGoldenrod}

\rule{\linewidth}{1pt}} 

\pretitle{\vspace{-80pt} \begin{flushleft} \HorRule \fontsize{14}{14}
\color{DarkRed} \selectfont}

\title{Simulation of Stochastic Non-Equilibrium Thermal Effects of 
Particle Inclusions within  
Fluid Interfaces and Membranes}

\posttitle{\par\end{flushleft}} 

\preauthor{\vspace{-15pt} \begin{flushleft}\fontsize{12}{12} 
\color{DarkRed}} 
\author{D. Jasuja$^{1}$, 
P. J. Atzberger$^{2}$ } 
\postauthor{\small \color{Black} 
\\ $[1]$ Department of Mathematics, University of California Santa Barbara (UCSB), \\
$[2]$ Department of Mathematics, Department of Mechanical Engineering, 
University of California Santa Barbara (UCSB); atzberg@gmail.com;
\url{http://atzberger.org/} \\
\vskip 1.5em

\vspace{-0.6cm}
\HorRule
\end{flushleft} 
} 

\DeclareGraphicsExtensions{.png}

\begin{document}
\date{}
\maketitle 

\thispagestyle{fancy} 

\vspace{-1.75cm}

\initial{W}\textbf{e formulate theoretical modeling approaches 
and develop practical computational
simulation methods for investigating the 
non-equilibrium statistical mechanics of fluid 
interfaces with passive and active immersed 
particles.  Our approaches capture phenomena
taking into account thermal and dissipative 
energy exchanges, hydrodynamic coupling, and 
correlated spontaneous fluctuations.  Our methods 
allow for modeling non-uniform time-varying 
temperature fields, fluid momentum fields, 
and their impacts on particle drift-diffusion dynamics.
We show how practical stochastic numerical methods  
can be developed for these systems by performing analysis to 
factor operators analytically to obtain efficient algorithms for 
generating the fluctuating fields.  We demonstrate
our methods in a few simulation studies showing how 
they can be used to capture particle heating 
of the interface and the roles of thermal gradients 
on hydrodynamic fluctuations. 
The methods developed provide modeling and simulation 
approaches that can be used for further 
investigations of
non-equilibrium phenomena in active soft materials, 
complex fluids, and biophysical systems.
}

\setlength{\parindent}{5ex}

\section*{Introduction}
Many soft materials and biophysical systems exhibit phenomena involving passive
and active energy exchanges in regimes out of thermodynamic equilibrium.
Examples include the viscoelastic responses of soft materials and complex
fluids~\cite{Bird1987Vol1,Bird1987Vol2,Doi2013,chen2010rheology,graham2018microhydrodynamics}, activity in biological cell transport
and
mechanics~\cite{Fang2019,Garcia2005,Howard2019,Rajagopal2019,Suzuki2020,
Talbot2017,mackintosh2010active,alberts2015essential,AtzbergerSoftMatter2016},
and experimental assays and processing techniques using thermal effects to manipulate colloids and
other small particles~\cite{Jiang2010,Duhr2006,Chen2021,jerabek2014microscale,piazza2008thermophoresis}.  Investigating
phenomena in these systems requires taking into account energy exchanges
arising from dissipative mechanisms in the mechanics, gradients in temperature,
and roles played by fluctuations.  This includes exchanges from
dissipation during coupling in hydrodynamic flows and temperature gradients
arising from particle chemical 
activity or external heating ~\cite{alberts2015essential,Chen2021,Duhr2006,
AtzbergerSoftMatter2016,MacKintosh2007nonequilibrium}.

Capturing these effects in a tractable and consistent manner poses
significant challenges for modeling and simulation.  This requires
stochastic spatial-temporal models accounting for both the mechanical
and thermodynamic exchanges along with appropriate fluctuations.
Most coarse-grained modeling and simulation approaches treat
fluctuations in a limiting regime where temperatures have
equilibrated to a global constant value.  
For soft materials and complex fluids, this includes 
simulation approaches such as Brownian-Stokesian 
Dynamics~\cite{Banchio2003,McCammon1988}, Coarse-grained 
approaches with Langevin thermostats~\cite{Doi2013,
Deserno2009multiscale,noid2013perspective},
and continuum mechanics formulations such as Stochastic Immersed
Boundary Methods
(SIBMs)~\cite{Atzberger2007,AtzbergerSoftMatter2016,AtzbergerSurfFluctHydro2022}
and Stochastic Eulerian Lagrangian Methods
(SELMs)~\cite{AtzbergerSELM2011,AtzbergerTabak2015}.  These computational
simulation methods are based on statistical mechanics primarily 
in regimes in thermodynamic equilibrium.

For investigations of soft materials with active microstructural
elements, biophysical systems, or experimental assays having external
sources of energy, such as laser heating of particles, requires
further theoretical modeling and simulation approaches to capture the
relevant roles of thermal effects and other energy exchanges.  For
this purpose, we develop here theoretical formulations and practical
computational methods for non-equilibrium thermodynamic regimes.  We build on
our past work on equilibrium and non-equilibrium statistical
mechanics and on Stochastic Eulerian Lagrangian Methods
(SELMs)~\cite{AtzbergerSELM2011,AtzbergerTabak2015,
AtzbergerSurfFluctHydro2022,AtzbergerSoftMatter2016,
Atzberger2013,AtzbergerNotes2021}.
As a specific case to demonstrate these approaches, we focus on the context of
fluid interfaces / membranes having particle inclusions.  We formulate our models to
account for non-uniform temperature fields, hydrodynamic coupling, heat
exchanges between the particles and fluid, and the roles of
fluctuations in the hydrodynamics and particle drift-diffusion
dynamics.  We develop SELM approaches for non-equilibrium
regimes for investigating fluid interfaces with  
active and passive particles.   Part of the motivations
for this work include the roles of activity of proteins in cellular
processes~\cite{MacKintosh2007nonequilibrium,mackintosh2010active,
mogilner2018intracellular,nematbakhsh2017multi,Alberts2002a,howard2019cytoplasmic,
suzuki2020challenge}, 
active inclusions within lipid bilayer 
membranes~\cite{AtzbergerSoftMatter2016,rajagopal2019transient,
safran2018statistical,AtzbergerSurfaceFluctHydro2022}, 
recent experimental systems that use
lasers to thermally excite and manipulate particles
\cite{Jiang2010,Duhr2006,Chen2021,jerabek2014microscale,piazza2008thermophoresis}, 
and other modeling and simulation challenges for active
processes in soft materials and biophysical systems
\cite{demirel2007nonequilibrium,Fang2019,garner2013cell,DogicActive2012}.

\section*{Materials and methods}

\subsection*{Stochastic Non-Equilibrium Model for Thermal Effects and Fluctuations}

We first formulate our theoretical modeling approach for capturing the 
mechanics and thermal effects within fluid-particle systems.  
We then discuss how practical stochastic 
numerical methods can be developed for performing simulations.  
We then perform a few simulations of specific models
to demonstrate the approaches.  

\begin{figure}[h]

\centerline{\includegraphics[width=0.95\columnwidth]{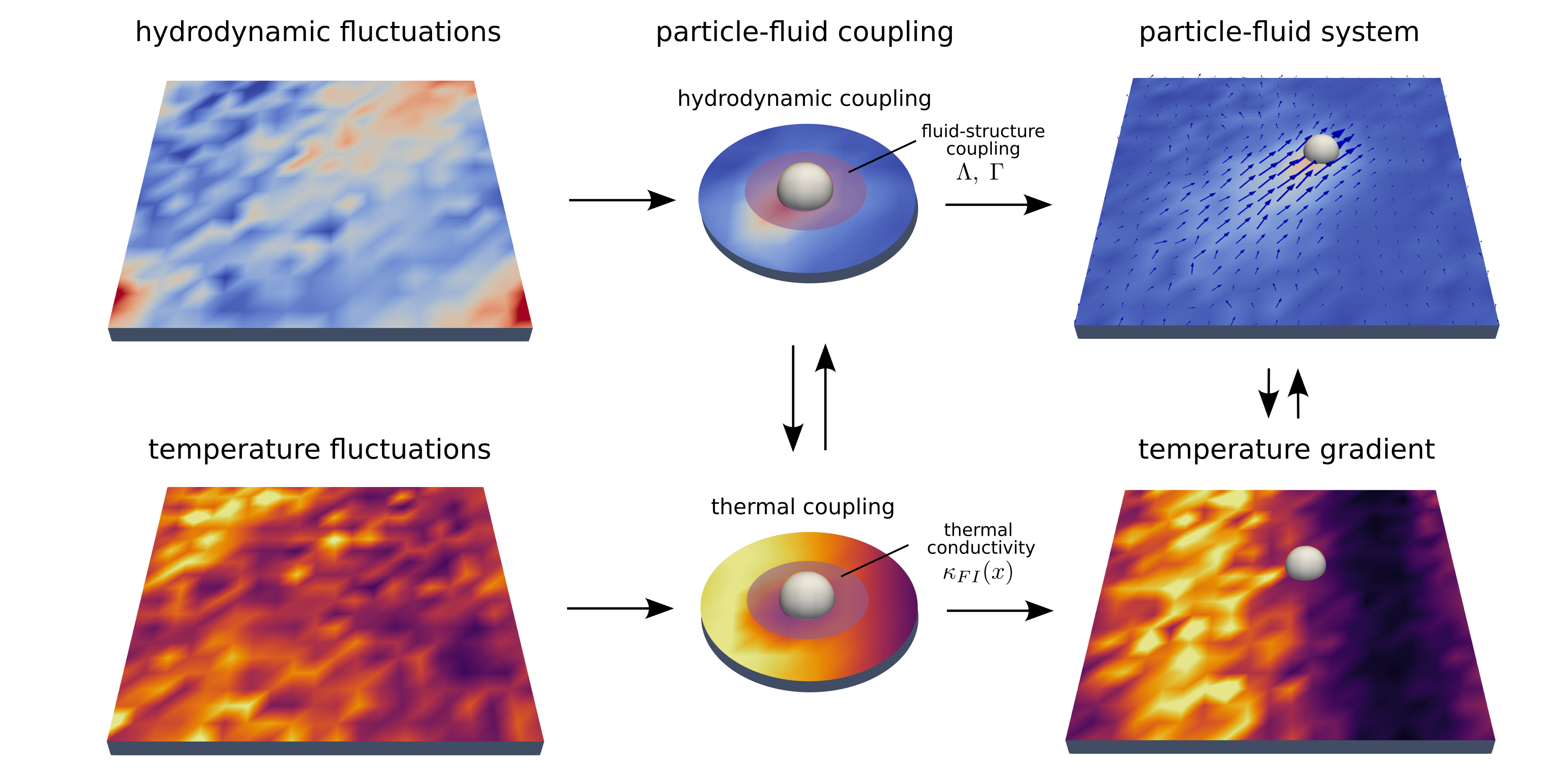}}
\caption{{\bf Stochastic Non-Equilibrium Model for the Particle-Fluid Interface.}
The model accounts for energy exchanges and fluctuations in the particle-fluid 
system from reversible and irreversible processes.  This includes exchanges between
the particle and fluid for non-uniform temperature fields, hydrodynamic flows, 
and other dissipative mechanisms.  The 
fluid-particle coupling is handled through tractable approximations to the 
fluid-structure interactions based on the averaging operators 
$\Lambda,\Gamma$ and for thermal exchanges through a spatially 
varying thermal conductivity 
$\kappa_{F,I}(x;X)$.  The approaches allow for capturing phenomena 
in non-equilibrium regimes arising from temperature gradients, heat exchanges, 
and fluctuations, see equation~\ref{equ_full_model}}
\label{fig_full_model}
\end{figure}

We model the system
using the following stochastic 
fluctuating hydrodynamics description coupled to a collection of
thermodynamic energy exchange equations of the form
\begin{eqnarray}
\label{equ_full_model}
\frac{d\mb{X}}{dt} & = & \mb{V} ,\;\hspace{0.4cm}
m \frac{d\mb{V}}{dt} = -\gamma\left(\mb{V} - \Gamma \mb{u}\right)
-\Xi{\mb{V}} -\nabla_{\mb{X}} \Phi(\mb{X}) + \mb{G}_{\tiny thm,p_P}  \\
\nonumber
\rho \frac{\partial \mb{u} }{\partial t} &=& \nabla \cdot \bsy{\sigma} + \bsy{\tau} -
\nabla p + \Lambda\left[ \gamma\left(\mb{V} - \Gamma \mb{u}\right)  \right]+
\mb{g}_{\tiny thm,p_F}, \hspace{0.4cm}
\nabla \cdot \mb{u} = 0 \\
\frac{\partial \theta_P}{\partial t} & = & -\frac{\kappa_{PI}\left(\theta_P -
\theta_I\right)}{c_P} 
+
\frac{\dot{q}_P^T D_P(Y)\dot{q}_P}{c_P} 
+ \mb{G}_{\tiny thm, \theta_P} \\
\nonumber
\frac{\partial \theta_F(x,t)}{\partial t} & = & 
\frac{\nabla\cdot \left(\kappa_{FF}\nabla \theta_F(x)\right)}{c_F}
-\frac{\kappa_{FI}(x;X)
\left(\theta_F - \theta_I \right)}{c_F}  \\
\nonumber
&+& \frac{\dot{q}_{F}^T D_F(Y)\dot{q}_{F}}{c_F} 
+ \mb{g}_{\tiny thm, \theta_F}(x,t) \\
\nonumber
\frac{\partial \theta_I}{\partial t} & = & \frac{\kappa_{PI}\left(\theta_P -
\theta_I\right)}{c_I} + \int \frac{\kappa_{FI}(x;X)\left( \theta_F - \theta_I
\right)}{c_I} dx +
\frac{\dot{q}_{P,F}^T D_I(Y)\dot{q}_{P,F}}{c_I} 
+ \mb{G}_{\tiny thm, \theta_I}.
\end{eqnarray}
The first part of our model employs momentum equations for the particle
drift-diffusion $X(t),V(t)$ coupled to fluctuating
hydrodynamic~\cite{Atzberger2007,AtzbergerSurfaceFluctHydro2022,AtzbergerSoftMatter2016}
equations for $u(x,t)$.  For tractable computations, we treat
the fluid-structure interactions using a coarse-grained interaction model 
having similarities to Immersed Boundary Methods
(IBMs)~\cite{Peskin2002,Atzberger2007} and Stochastic Eulerian Lagrangian Methods
(SELMs)~\cite{AtzbergerSELM2011,AtzbergerTabak2015}.  More details
are discussed below.  The second part of the equations accounts for the irreversible
dissipative energy exchanges within the system which generate heat and direct
heat exchanges that impact the temperature fields.  We model our system with
three types of heat bodies (i) particles with temperature $\theta_P$, (ii)
fluid interface with temperature field $\theta_F(x)$, and (iii) interfacial
region involved in the fluid-structure interactions with temperature
$\theta_I$.  The $\mb{G}_{thm,*}, \mb{g}_{thm,*}$ are stochastic terms
accounting for fluctuations. 
We denote the vector-valued terms as $\mb{G}_{thm}(t)$
and the 
stochastic spatial fields as
$\mb{g}_{thm}(x,t)$.  A schematic of the
model can be found in Figure~\ref{fig_full_model}.

We now give some more details. For this purpose, it is convenient to collect
together all of the degrees of freedom of the system by letting  
\begin{eqnarray}
\label{equ_Y_def}
Y = 
\left[
\mb{X}, 
\bsy{\phi},
\mb{R},
m\mb{V},
\rho \mb{u},
\tilde{m}\dot{\mb{R}},
\theta_P,
\theta_F,
\theta_I
\right]^T = [\mb{q},\mb{p},\bsy{\theta}]^T,
\end{eqnarray}
where $\mb{q} = [q_P,q_F,q_I] = [\mb{X},\bsy{\phi},\mb{R}]$ 
and $\mb{p} = [p_P,p_F,p_I] = [m\mb{V},\rho\mb{u},\tilde{m}\dot{\mb{R}}]$,
and $\bsy{\theta} = [\theta_P,\theta_F,\theta_I]$.  
For completeness and to allow for 
simplified common expressions within 
our derivations, we include a few
additional degrees of freedom.  These are
$\bsy{\phi}$ for the fluid displacement map,
$\mb{R}$ the interface state vector, 
$\tilde{m}\dot{\mb{R}}$ the interface 
pseudo-momentum.  In practice, these do not play a significant role in our
current model but are helpful to include so terms can be treated similarly
in our derivations.  

The fluid-structure interactions 
are treated by a drag force in the particle equations referencing the local flow 
environment and an equal-and-opposite body-force term in the fluid equations.
For this purpose, the $\Gamma[u] = \int \delta_a(x - \mb{X}) u(x,t) dx$ 
operator is used to probe the local flow environment to determine
a reference velocity at the particle location $\mb{X}$.  The $\Lambda$ operator
is used to distribute forces spatially to a region of the fluid nearby to
$\mb{X}$.  For this purpose we use the adjoint operator 
$\Lambda = \Gamma^T$ with $\Lambda[F] = F\delta_a(x - X)$.
Many other choices can be made for the forms of $\Gamma,\Lambda$, we 
use throughout the above with $\delta_a$ the 
Peskin $\delta$-function with
$a=\Delta{x}$~\cite{Atzberger2007,Peskin2002,AtzbergerSoftMatter2016}.
The $\gamma$ provides the strength of the drag coupling.  The $\Xi$
is a symmetric positive semi-definite operator accounting for any friction
effects treated as directly between the particles.  
The $\Phi(\mb{X})$ provide a potential energy resulting in
forces acting on the particles. 

The in-plane hydrodynamic stress is denoted
by $\bsy{\sigma}$ and the traction stress with the surrounding medium 
is denoted by $\bsy{\tau}$.
We use a Newtonian in-plane stress 
$\bsy{\sigma} = \mu\left(\nabla\mb{u} +
\nabla \mb{u}^T\right)$ with $\mu$ the fluid viscosity and $p$ the
fluid pressure.  
In this initial work, we use for simplicity the traction stress
$\bsy{\tau} = -\lambda u$ with the dissipative viscosity $\lambda$. 
In our model and numerical methods, alternative in-plane and traction 
stresses also can be treated readily in the methods to 
use other 
hydrodynamic models and theories arising for 
thin films and membranes~\cite{AtzbergerSoftMatter2016,
AtzbergerSurfaceFluctHydro2022,Saffman1975,Saffman1976}.

The temperatures $\theta_P,\theta_F(x),\theta_I$ 
account for energy exchanges from dissipation in
the mechanics and direct heat exchanges.  Associated
with these temperatures are three types of heat bodies.
These are related to different
dissipative mechanisms reflecting the rates at which
the system does work in the form of friction. For the 
fluid-particle system these are

\begin{eqnarray}
\label{equ_D_def}
\dot{q}_P^TD_P(Y)\dot{q}_P &=& V^T\Xi V \\
\dot{q}_F^T D_F(Y) \dot{q}_F &=& 
\mu \nabla u : \left(\nabla u + \nabla u^T\right) \\
\dot{q}_{P,F}^TD_{I}(Y)\dot{q}_{P,F} &=& 
V^T \gamma \left( V - \Gamma u\right)  
-u^T \gamma \Lambda\left[V - \Gamma u\right].
\end{eqnarray}
The $\dot{q}$ gives the rates of change of the mechanical 
configuration.  We distinguish the different parts of the 
system using the notation  
$\dot{q}_F = u$, $\dot{q}_P = V$, and $\dot{q}_{P,F} = [V,u]^T$.
We remark for simplicity and brevity in this initial work, 
we handle the $-\lambda u$ by treating the coupling
and $\lambda$ as small having a negligible effect on 
the surrounding medium and the system.

In practice,
depending on the system and regime this may require 
further consideration and more detailed treatment 
of the surrounding medium.

We briefly discuss the thermodynamics of our system associated with 
these energy exchanges
and fluctuations.  As a basic model we use the 
following internal energies and entropies. 
For the particles and interfacial region
we formulate entropies as
$\mathcal{S}^{(1)} = s_1(\theta_P) = c_P\ln(\theta_P)$, $\mathcal{S}^{(2)} =
\int s_2(\theta_F) dx = \int c_F\ln(\theta_F(x))dx $,
$\mathcal{S}^{(3)} = s_3(\theta_I) = c_I\ln(\theta_I)$.
The $c_P,c_F,c_I$ denote the specific heats.   
The fluid body is spatially extended and 
we emphasize that this is formulated using the local entropy densities 
$s_2(x,\theta_F(x)) = c_F \ln(\theta_F(x))$.  This gives the system total entropy 
$\mathcal{S} = \mathcal{S}^{(1)} + \mathcal{S}^{(2)} + \mathcal{S}^{(3)}$.
Similarly, we have for the particles and interfacial region 
internal energies $\mathcal{U}^{(1)} = u_1(\theta_P) = c_P\theta_P$,
$\mathcal{U}^{(3)} = u_3(\theta_I) = c_I\theta_I$.  Since the fluid body is spatially 
extended, we formulate using the local internal energies
$u_2(x,\theta_F(x)) = c_F \theta_F(x)$ giving the global energy $\mathcal{U}^{(2)} = \int
u_2(x,\theta_F(x)) dx = \int c_F \theta_F(x) dx$.  This gives for the system the total energy
\begin{eqnarray}
\mathcal{E} = \frac{1}{2m} p_P^2 + \Phi(q_P) + \int \frac{1}{2\rho} p_F(x) dx
+ \frac{1}{2\tilde{m}} p_I^2 
+ \mathcal{U}^{(1)} +\mathcal{U}^{(2)} +
\mathcal{U}^{(3)}. 
\end{eqnarray}
The terms involving the momentum $p_P,p_F,p_I$ give the kinetic
energy of the system. The $\Phi(q_P)$ is the energy associated with
the particle potential energy in equation~\ref{equ_full_model}.

In the context of soft materials and biophysical systems, 
we are often interested in phenomena in regimes having 
spatial-temporal scales where diffusive effects play an important role.  This
requires a non-trivial extension of the above formulation to account for the
entropy production from dissipation and spontaneous energy exchanges from thermal fluctuations.
As part of this modeling, we will use as a guideline the abstract framework of
non-equilibrium statistical mechanics referred to as
\textit{GENERIC}~\cite{Ottinger1997,Ottinger1997b}.  This framework generalizes
Ginzsburg-Landau
theory~\cite{Ottinger1997,Ottinger2005,Landau2013,Mielke2011,grmela2018generic,
Hohenberg2015}
and requires that the reversible and irreversible processes be formulated in
terms of an anti-symmetric operator $L$ and a positive semi-definite symmetric
operator $K$~\cite{Ottinger1997,Ottinger1997b,Ottinger2005,
grmela2018generic,vazquez2005fluctuation}.  
These act respectively on the energy gradient $\nabla E$ and
entropy gradient $\nabla S$ to obtain dynamics that have the abstract form
$dY_t = L \nabla Edt + K \nabla S dt + k_B(\nabla \cdot K)dt + BdW_t$.  The
fluctuations satisfy $BB^T = 2k_BK$, where $k_B$ is the Boltzmann constant.  We
will also build on our prior work for equilibrium and non-equilibrium 
fluctuations for SELMs and other systems with uniform and 
non-uniform stochastic fields~\cite{Atzberger2007,
AtzbergerTabak2015,AtzbergerRD2010,AtzbergerSELM2011,AtzbergerNotes2021}.

Given the three types of heat-bodies involved in the energy exchanges, our
prior derivations shows it is convenient to treat the system using a collection 
of operators $K^{(j)}$ each corresponding to one of the heat bodies.  
This gives dynamics of the general abstract form 
\begin{eqnarray}
\label{equ_gen_stoch_form}
dY_t = L \nabla Edt + \sum_j K^{(j)}
\nabla S^{(j)} dt + k_B(\nabla \cdot K^{(j)})dt + B^{(j)}dW_t^{(j)}.
\end{eqnarray}
In this formulation, the 
fluctuations satisfy $B^{(j)}B^{(j),T} = 2k_BK^{(j)}$.  The $S^{(j)}$ denotes
the entropy of the $j^{th}$ heat body. 
For the particle-fluid interface
system this yields the stochastic dynamics incorporating both the temperature
fluctuations and momentum fluctuations of the particle-fluid system.

To obtain more specific forms for the stochastic fluctuations
$\mb{G}_{thm}, \mb{g}_{thm}$, we perform 
derivations below in the context of development of our stochastic 
numerical methods.  We remark that to obtain practical simulation methods for 
equation~\ref{equ_full_model} requires we develop numerical 
approaches for approximating both the differential operators and 
stochastic terms $\mb{G}_{thm}, \mb{g}_{thm}$.  
This can be done most conveniently by reformulating
our model in equation~\ref{equ_full_model} 
in terms of the above more abstract framework in 
equation~\ref{equ_gen_stoch_form}.  This
requires we identify the operators $L$, $K^{(j)}$ corresponding to 
equation~\ref{equ_full_model} and perform some additional
analysis.  In numerical methods for the stochastic terms, 
we need to be able 
compute the stochastic terms having statistics equivalent to
$\zeta = \sqrt{2k_B} B^{(j)}\xi$  where $\xi$ are Gaussians with 
$\langle \xi \xi^T \rangle = I$, and
$\langle \zeta \zeta^T \rangle = 2k_B  B^{(j)}B^{(j),T}  =  2k_B K^{(j)}$.
While in principle, Choleksy factorizations could be used for this
purpose~\cite{strang1986introduction,faires2012numerical}, 
given that $B^{(j)}(Y)$ depends on the state of the 
system this computation would be required each time-step incurring
a cost of $O(n^3)$. The $n$ is the number of rows of 
$K^{(j)}$, which can be large especially for the 
discretized fluid fields.  For this purpose, we will  
derive analytically factorizations $B^{(j)}$ for $K^{(j)}$
allowing for more efficient generation of the stochastic driving
fields for performing simulations. 

\section*{Stochastic Numerical Methods} 
We now discuss briefly the stochastic numerical methods and fluctuations in our model in 
equation~\ref{equ_full_model}.  To provide a unified approach to deriving the 
fluctuations and our numerical discretizations, we 
reformulate our system in terms of stochastic dynamics of the general form 
\begin{eqnarray}
\label{equ_gen_stoch_form_2}
dY_t = L \nabla Edt + \sum_j K^{(j)} \nabla S^{(j)} dt + k_B(\nabla \cdot K^{(j)})dt + B^{(j)}dW_t^{(j)},
\end{eqnarray}
where $B^{(j)}B^{(j),T}  =  2k_B K^{(j)}$.  This stochastic differential equation is to be given an Ito
interpretation~\cite{Oksendal2000}. 
The $Y$ denotes the state of our fluid-particle system with components 
\begin{eqnarray}
\label{equ_Y_def}
Y = 
\left[
\mb{X}, 
\bsy{\phi},
\mb{R},
m\mb{V},
\rho \mb{u},
\tilde{m}\dot{\mb{R}},
\theta_P,
\theta_F,
\theta_I
\right]^T = [\mb{q},\mb{p},\bsy{\theta}]^T.
\end{eqnarray}
This can be expressed as $\mb{q} = [q_P,q_F,q_I] = [\mb{X},\bsy{\phi},\mb{R}]$ 
and $\mb{p} = [p_P,p_F,p_I] = [m\mb{V},\rho\mb{u},\tilde{m}\dot{\mb{R}}]$,
and $\bsy{\theta} = [\theta_P,\theta_F,\theta_I]$.  
For completeness and for allowing for 
simplified common expressions within 
our derivations, we include a few
additional degrees of freedom.  These are
$\bsy{\phi}$ for the fluid displacement map,
$\mb{R}$ the interface state vector, 
$\tilde{m}\dot{\mb{R}}$ the interface 
pseudo-momentum.  In practice, these do not 
play a significant role in the current stochastic dynamics
allowing us to avoid tracking them in our implementations,
but they do provide convenience for making a 
common treatment of expressions in our derivations. 

We start in this initial work with the basic stochastic time-step integrator
\begin{eqnarray}
Y^{n+1} &=& Y^n + L(Y^n)\nabla E(Y^n) \Delta{t} + \sum_j K^{(j)}(Y^n) \nabla S^{(j)}(Y^n) \Delta{t} \\
\nonumber
&+& \sum_j \left(\nabla \cdot K^{(j)}(Y^n)  \right) \Delta{t} + \mb{g}^n.
\end{eqnarray}
This is based on the Euler-Maruyama approach~\cite{Platen1992}.  The $Y^n = Y(t_n)$, $\Delta{t}$ is
the time-step, and $\mb{g}^n$ denotes the stochastic forcing with $\langle \mb{g}^{n_1} \mb{g}^{n_2} \rangle 
= \delta_{n_1,n_2} 2 k_B \sum_j K^{(j)}\Delta{t}$.
Other time-step integrators preserving additional structures also can be developed readily
based on stochastic Verlet, Runge-Kutta, or Multistep 
schemes~\cite{Platen1992}.  This inital work focuses primarily on
the spatial discretizations, generation of fluctuations, and related 
numerical considerations.   

For brevity, and to help make clear the structures
that need to be preserved in our discretizations, we present at the same
time our derivations for the stochastic driving fields
for the fluctuations and for our numerical methods.  An important
underlying structure in these derivations arises from the properties of 
the gradient operator $\nabla$ and 
divergence operator $\nabla \cdot = \mbox{div}$, and their 
negative adjoint relationship $\nabla \cdot = \mbox{div} = -\nabla^T$. 
As such, we retain the notation $\nabla$ and $\nabla \cdot$ for our 
operators in our derivations and in our numerical methods
replace these with numerical discrete operators 
with $\mathcal{G} \sim
\nabla$ and $\mathcal{D} \sim \mbox{div} = \nabla \cdot$,
while imposing that $\mathcal{D} = -\mathcal{G}^T$.  We give below a finite volume
discretizations for $\mathcal{D}$ and $\mathcal{G}$ 
satisfying these conditions after we present the general
derivations.  We then
use these derivations to generate our numerical methods by replacing $\nabla$
and $\nabla \cdot = \mbox{div}$ with $\mathcal{G}$ and $\mathcal{D}$ throughout 
in our analytic expressions.  This approach provides a
systematic way to obtain discretizations and stochastic numerical methods 
preserving important structures of the dynamics helping to ensure 
they yield reliable physical results.

\subsection*{Operators of the Reversible and Irreversible Processes}
This formulation of our dynamics uses operators that act on
the energy gradient $\nabla E$ and entropy gradient 
$\nabla S$.  Relative to the continuum formulation 
there is a subtle but important distinction in our 
numerical discretizations.
We discretize our system using a 
finite volume perspective, where we replace spatial integrals 
$\int (\cdot) dx$ by the corresponding finite sum 
$\sum_{m} (\cdot)_{x_m} \dvol$.  We treat continuum bodies as 
divided into a discrete finite collection of boxes each having 
volume $\dvol$. The $x_m$ denotes the location
of the $m^{th}$ finite volume box.  We use for 
our total entropy gradient and energy gradients
\begin{eqnarray}
\label{equ_grad_S}
\label{equ_grad_E}
\nabla \mathcal{S} 
&=& 
\left[
\partial_{\mb{q}} \mathcal{S},
\partial_{\mb{p}} \mathcal{S},
\partial_{\bsy{\theta}} \mathcal{S}
\right]^T 
=
\left[
0,\ldots
0,
c_P/\theta_P,
c_F\dvol /\theta_F,
c_I/\theta_I
\right]^T \\
\nabla \mathcal{E} 
&=& 
\left[
\partial_{\mb{q}} \mathcal{E},
\partial_{\mb{p}} \mathcal{E},
\partial_{\bsy{\theta}} \mathcal{E}
\right]^T 
= 
\left[\nabla_{\mb{X}}\Phi(\mb{X}), 0,0,
\mb{V}, \mb{u}\dvol, \mb{\dot{R}},
c_P,c_F\dvol,c_I
\right]^T.
\end{eqnarray}
We further distinguish entropy gradient components depending on which
irreversible process is being considered.  For this purpose, we use the
notation 
\begin{eqnarray}
\nabla S^{(1)} &=& \nabla_{Y} S^{(1)}(\theta_P) = \left[
0,\ldots
0,
c_P/\theta_P,
0,
0
\right]^T \\
\nabla S^{(2)} &=& \nabla_{Y} S^{(2)}(\theta_F) = \left[
0,\ldots
0,
0,
c_F\dvol /\theta_F,
0
\right]^T  \\
\nabla S^{(3)} &=& \nabla_{Y} S^{(3)}(\theta_P,\theta_F,\theta_I) = \left[
0,\ldots
0,
c_P/\theta_P,
c_F\dvol /\theta_F,
c_I/\theta_I
\right]^T. 
\end{eqnarray}
The extra factor of $\dvol$ arises in these gradients from
the energy and entropy density contributions.  In the 
energy setting the kinetic energy density now 
contributes in the finite volume discretization
for each box as
$\frac{1}{2}\frac{p_F^2}{\rho} \dvol = \frac{1}{2}\frac{(\rho u)^2}{\rho} \dvol$.  
Since our degrees of freedom are $p_F(x_m) = \rho u(x_m)$ for the $m^{th}$ box, we
now also have an additional factor of $\dvol$.  This similarly holds for the 
entropy gradients.  This will also 
contribute to our discretizations for the operators.

Using these gradients, we next reformulate our model in
terms of the operators $L$ and $K^{(j)}$.  The reversible parts of
the dynamics can be reformulated similar to Hamiltonian mechanics 
to use the anti-symmetric operator \\
\begin{eqnarray}
\label{equ_L_def}
L 
= 
\begin{bNiceMatrix}[last-row=5,last-col=5]
0  & 0 & I        & 0  & \atzmlabel{$q_P$} \\
0  & 0 & 0        & I/\dvol & \atzmlabel{$q_F$}  \\
-I & 0 & 0        & 0 & \atzmlabel{$p_P$} \\
0  & -I/\dvol  & 0 & 0 & \atzmlabel{$p_F$} \\
\atzmlabel{$q_P$} & \atzmlabel{$q_F$} & \atzmlabel{$p_P$} & \atzmlabel{$p_F$} & 
\end{bNiceMatrix}.
\end{eqnarray}
For brevity in our notation for the operators, we show just the non-zero subset
of the rows and columns of the operators.  For $L$ this corresponds to the
degrees of freedom $q_P,q_F,p_P,p_F$ for the fluid and particles, as defined
above.  The matrix representation for the operator acting for the full system
$\nabla_Y E$ can be obtained straight-forwardly by padding the other rows and
columns with zeros.  

We can reformulate the irreversible parts of our dynamics in terms of the
operators $K^{(j)}$.  For the dissipation in the mechanics of the particles we
have 
\begin{eqnarray}
\label{equ_K_1_def}
K^{(1)} 
= 
\begin{bNiceMatrix}[last-row=3,last-col=3]
\theta_PD_P(Y) & 
-\frac{\theta_P D_P (Y)\dot{q}_P}{c_P} & \atzmlabel{$p_P$} \\
-\frac{\dot{q}_P^T\theta_PD_P(Y)}{c_P} & \frac{\dot{q}_P^T\theta_PD_P(Y)\dot{q}_P}{c_P c_P} & \atzmlabel{$\theta_P$}  \\
\atzmlabel{$p_P$} & \atzmlabel{$\theta_P$} & 
\end{bNiceMatrix}.
\end{eqnarray}
where $D_P$ is given in equation~\ref{equ_D_def}.  This can be verified by
considering the action of this operator on the gradient of the entropy
$K^{(1)}\nabla S^{(1)}$ giving the contributions to the dynamics in
equation~\ref{equ_full_model}, and by considering the net energy exchanges in
the system which from our accounting of both mechanical and internal energy
requires $K^{(1)}\nabla_{Y^{(1)}}E = 0$.  Similarly, we can reformulate our
dynamics for the dissipation in the hydrodynamics to obtain the operator 
\begin{eqnarray}
\label{equ_K_2_def}
K^{(2)} 
= 
\begin{bNiceMatrix}[last-row=3,last-col=3]
-\frac{\wp\mbox{\small div}\left(\mu \theta_F \left(\nabla + \nabla^T \right) \wp^T \right)}{\dvol} & 
\frac{\wp\mbox{\small div}\left(\square \mu \theta_F \left(\nabla u + \nabla u^T \right) \right)}{c_F \dvol} & \atzmlabel{$p_F$} \\
\frac{-\nabla u:\left(\mu\theta_F\left( \nabla + \nabla^T \right)\wp^T\right) }{c_F \dvol }  & 
\frac{\nabla u:\left( \square \mu \theta_F \left(\nabla u + \nabla u^T\right)\right)}
{c_F c_F \dvol } + \frac{-\nabla \cdot \left(\tilde{\kappa}_0 \nabla \right)}{c_F c_F \dvol }& \atzmlabel{$\theta_F$} \\
\atzmlabel{$p_F$}  & \atzmlabel{$\theta_F$} &  \\
\end{bNiceMatrix}.
\end{eqnarray}
We take here $\tilde{\kappa}(\theta_F) =
\kappa_0 \theta_F^2$. 
The $\square$ in this notation indicates the location in the expression where
to insert the terms on which this operator acts.  The $\wp$ denotes the
projection operator associated with the incompressibility constraint on the
fluid~\cite{Atzberger2007,AtzbergerTabak2015}. We further use its adjoint
$\wp^T$ and that the operator is self-adjoint $\wp^T = \wp$.  We also use that
from the dynamics that $\wp u = u$ throughout.   We remark that in the discrete
case, we would replace the integrals $\int (\cdot) dx$ 
by the finite volume sums of the form
$\sum_m (\cdot)_{x_m} \dvol$.  For clarity and to avoid clutter in
expressions, we retain the integral notation throughout our derivations.

For the particle-fluid interface dissipation and direct
heat exchanges we have
{\fontsize{6}{8} \selectfont
\begin{eqnarray}
\label{equ_K_3_def}
K^{(3)} 
= 
\begin{bNiceMatrix}[last-row=5,last-col=5]
\theta_ID_I(Y) & 0 & 0 & \frac{-\theta_ID_I(Y)\dot{q}_{P,F}}{c_I} & \atzmlabel{$p_{P,F}$} \\
0 & \frac{\kappa_{PI}\theta_I\theta_P}{c_{P,P}}& 0 & -\frac{\kappa_{PI}\theta_P\theta_I}{c_{P,I}} & \atzmlabel{$\theta_P$} \\
0 & 0 &  \frac{\mbox{\tiny diag}(\kappa_{FI}\dvol\;\theta_F\theta_I)}{c_{F,F} \dvol \dvol} 
& -\frac{\kappa_{FI}\dvol\;\theta_I\theta_F}{c_{F,I} \dvol} & \atzmlabel{$\theta_F$} \\
\frac{-\dot{q}^T\theta_ID_I(Y)}{c_I} & 
-\frac{\kappa_{PI}\theta_I\theta_P}{c_{I,P}} & 
-\frac{(\kappa_{FI}\dvol\;\theta_I\theta_F)^T}{c_{I,F} \dvol} & 
\frac{\dot{q}^T\theta_I D_I(Y)\dot{q}}{c_{I,I}} +
\frac{\kappa_{PI}\theta_P\theta_I + \theta_I \int \kappa_{FI} \theta_F dx}{c_{I,I} } & \atzmlabel{$\theta_I$} \\
\atzmlabel{$p_{P,F}$} 
& \atzmlabel{$\theta_P$} 
& \atzmlabel{$\theta_F$} 
& \atzmlabel{$\theta_I$} & \\
\end{bNiceMatrix}.
\end{eqnarray}
}
We use notation $c_{ij} = c_i\cdot c_j$ for brevity.  
We now give the divergence $\nabla \cdot K^{(j)}$ for each of these terms.
For the particle dissipation we have  
\begin{eqnarray}
\label{equ_div_K_1}
\nabla_{Y^{(1)}} \cdot K^{(1)} =  
\begin{bNiceMatrix}[last-col=2]
-\frac{D_P(Y)\dot{q}_P}{c_P} & \atzmlabel{$p_P$}  \\
\frac{-m^{-1}\theta_P\mbox{tr}(D_P(Y))}{c_P} +
\frac{\dot{q}_P^TD_P(Y)\dot{q}_P}{c_P c_P} & \atzmlabel{$\theta_P$}.  \\ 
\end{bNiceMatrix}
\end{eqnarray}
where $Y^{(1)} = [p_P,\theta_P]^T$. In the notation to help keep expressions
compact, we only show the non-zero rows of the divergence vector.  Similarly,
we highlight this through $Y^{(1)}$ showing the only non-zero terms we need to 
consider when taking the derivatives.  This vector can then be expanded to
the full set of degrees of freedom by padding with zeros in the stochastic
dynamics in equation~\ref{equ_gen_stoch_form_2}.  
For the fluid dissipation, we have the divergence 
\begin{eqnarray}
\label{equ_div_K_2}
\nabla_{Y^{(2)}} \cdot K^{(2)} =  
\begin{bNiceMatrix}[last-col=2]
\frac{\wp \mbox{\small div}\left(\frac{\partial \mu\theta_F\left(\nabla u + \nabla u^T \right)}{\partial \theta_F} \right)}{c_F} & \atzmlabel{$p_F$} \\
\frac{-\rho^{-1}\nabla:\left(\mu\theta_F\left(\nabla + \nabla^T\right) \wp^T \right)}{c_F} +
\frac{\nabla u:\frac{\partial \mu\theta_F\left(\nabla u + \nabla u^T \right)\wp^T}{\partial \theta_F}} 
{c_F c_F} 
+ 
\frac{
\frac{\partial -\nabla \cdot \left(\tilde{\kappa}(\theta_F) \nabla \right)}{\partial \theta_F}}{c_F c_F} & \atzmlabel{$\theta_F$} 
\end{bNiceMatrix}.
\end{eqnarray}
The $Y^{(2)} = [p_F(x),\theta_F(x)]^T$. 
For the divergence for the dissipation from the interfacial fluid-particle coupling and 
the direct heat exchanges between particles, fluid, and interface we 
have 
\begin{eqnarray}
\label{equ_div_K_3}
\nabla_{Y^{(3)}} \cdot K^{(3)} =  
\begin{bNiceMatrix}[last-col=2]
\frac{-D_I(Y)\dot{q}_{P,F}}{c_I} 
& \atzmlabel{$p_{P,F}$}  \\
\frac{\kappa_{PI}\theta_I}{c_{P,P}} -\frac{\kappa_{PI}\theta_P}{c_{P,I}}& \atzmlabel{$\theta_P$}  \\
\frac{\kappa_{FI}\dvol \;\theta_I}{c_{F,F} \dvol \dvol} -\frac{\kappa_{FI}\dvol\;\theta_F}{c_{F,I}\dvol} & \atzmlabel{$\theta_F$}  \\ 
\frac{-\theta_I\mbox{tr}\left(D_I(Y) M^{-1}_{P,F}\right)}{c_I}
+
\frac{\dot{q}^TD_I(Y)\dot{q}}{c_I c_I}
-\frac{\kappa_{PI}\theta_I}{c_{I,P}} 
-\frac{\int \kappa_{FI}dx\;\theta_I }{c_{I,F} \dvol}
+
\frac{\kappa_{PI}\theta_P + \int \kappa_{FI} \theta_F dx}{c_{I,I}}& \atzmlabel{$\theta_I$}  
\end{bNiceMatrix}
\end{eqnarray}
The $Y^{(3)} = [p_{P,F},\theta_P,\theta_F,\theta_I]^T$ with $p_{P,F} = [p_P,p_F]^T$
and $M^{-1}_{P,F} = \frac{d \dot{q}_{P,F}}{d p_{P,F}}$.

To obtain practical stochastic numerical methods requires we are 
able to generate the thermal fluctuations for each time-step.  We now
present some methods avoiding Cholesky factorizations
by performing analysis to obtain directly factorizations $B^{(j)}(Y)$ 
for each of the $K^{(j)}(Y)$.  

\subsection*{Methods for Generating the Thermal Fluctuations}
We now briefly discuss our approach for generating the needed 
stochastic fields for the  
thermal fluctuations with correlations $B^{(j)}$.  
We will perform analysis that yields direct factorizations of $K^{(j)}$. 
We establish a few identities useful in obtaining factors.  As one identity, we will use that for 
operators that can be decomposed into a sum of positive 
semi-definite operators $C = C_1 + C_2$ can be factored using
$C_1 = R_1R_1^T$ and $C_2 = R_2R_2^T$ to
generate the needed noise $\zeta$.  To obtain the covariance 
$\langle \zeta \zeta^T \rangle = C$ we use for our generation 
procedure 
$\zeta = \zeta_1 + \zeta_2$ with 
$\zeta_1 = R_1\xi_1$ and $\zeta_2 = R_2\xi_2$ 
generated independently
with $\xi_1 \sim \eta(0,I)$ and $\xi_2 \sim \eta(0,I)$.
The $\eta(0,I)$ denotes the distribution of vectors having standard Gaussian 
components yielding mean zero and covariance the identity matrix $I$.  
This gives the needed stochastic variates, since $\langle \zeta \zeta^T \rangle = \langle 
\zeta_1\zeta_1^T\rangle + \langle \zeta_2\zeta_2^T \rangle
= R_1R_1^T + R_2R_2^T = C_1 + C_2 = C$, using $\langle
\zeta_1 \zeta_2^T\rangle = 0$.  We show how this and 
other factorization techniques can be employed to obtain 
practical algorithms for generating the needed 
fluctuations for the spatial hydrodynamic fields,
particles, and thermal fields.  

We generate fluctuations for the particles $X(t),V(t)$ and their 
temperature $\theta_P$ by decomposing the operator as 
$K^{(1)} = RR^T$.  For this purpose, we use the factor
\begin{eqnarray}
\label{equ_R_K_1}
R = 
\begin{bNiceMatrix}[last-row=3,last-col=2]
\sqrt{\theta_P}R_D(Y) & \atzmlabel{$p_P$} \\
-\frac{\sqrt{\theta_P}\dot{\mb{q}}_P^TR_D(Y)}{c_P}& \atzmlabel{$\theta_P$} \\
\atzmlabel{$p_P$} \\
\end{bNiceMatrix}.
\end{eqnarray}
This requires we use $R_D$ so that 
$R_DR_D^T = D_P(Y)$, where $D_P$ is from equation~\ref{equ_D_def}. 
For the particles, this depends
on the form of the operator $D_P(Y) = \Xi$.  In the present work,
we take $\Xi = 0$, making this trivial.  More generally in the 
worst case a Cholesky factorization of just $\Xi$ can be utilized,
if a factorization is not readily available.  The expression
above gives a general procedure for leveraging knowledge of 
a factorization of $\Xi$ for use in our modeling and simulation 
framework.  We further remark that 
fluctuations in this case only require random variates of
the same dimension as $p_P$ and do not require independent
terms for the temperature.  Intuitively, this can be explained
since the temperature fluctuations are from the same  
spontaneous mechanism of energy exchange that impacts the 
particle momentum fluctuations.  This requires the specified
correlations to ensure the energy balances.  

We generate fluctuations for the hydrodynamics $u(x,t)$ and their 
temperature fields $\theta_F(x)$ by using 
the following factor
\begin{eqnarray}
\label{eqn_R0_fluid}
R & = & 
\begin{bNiceMatrix}[last-row=3,last-col=3]
-\mbox{\small div}\left(R_{visco}\right) & 0 & \atzmlabel{$p_F$} \\
\frac{-\nabla{u}:R_{visco}}{c_F} & \frac{R_{heat}}{c_F}& \atzmlabel{$\theta_F$} \\
\atzmlabel{$p_F$} 
& \atzmlabel{$\theta_F$} & \\
\end{bNiceMatrix}.
\end{eqnarray}
The $R_{visco}R_{visco}^T = K_{visco}$ and
$R_{heat}R_{heat}^T = K_{heat}$.
We define the operator $K_{visco} A = \frac{\mu \theta_F(x)}{\dvol} \left(A + A^T\right)$, which acts on a tensor to take the symmetric part and multiply
by the local temperature $\theta_F(x)$ and viscosity $\mu$.  We define $K_{heat} = -\mbox{\small div}\left( \tilde{\kappa} \nabla\right)$.

For factorization of $R_{visco}$, it is convenient to express the operator as 
$K_{visco} = \frac{\mu \theta}{\dvol} \mathcal{T}$ with $\mathcal{T}A = A + A^T$.  To obtain $R_{visco}$ we use the following properties of $\mathcal{T}$.
The operator is self-adjoint, since $\mathcal{T}A = A + A^T$ gives $\langle
\mathcal{T}A,B \rangle = \langle A + A^T,B \rangle = A:B + A^T:B$
and $\langle A,\mathcal{T}B \rangle = \langle A,B + B^T \rangle = A:B + A:B^T =
A:B + A^T:B = \langle \mathcal{T}A,B \rangle$.  Hence the operator $\mathcal{T}
= \mathcal{T}^T$ is self-adjoint and has a symmetric matrix representation.
We further have that 
$\mathcal{T}^2A = \mathcal{T}(A + A^T) = (A + A^T) + (A^T + A) = 2(A + A^T) =
2\mathcal{T}A$.  If we let $R = \frac{1}{\sqrt{2}} \mathcal{T}$ then $RR^T =
\frac{1}{2}\mathcal{T}\mathcal{T}^T = \frac{2}{2} \mathcal{T} = \mathcal{T}$.
This gives the needed factorization.   The $R_{visco}A = \sqrt{\frac{1}{2}
\frac{\mu \theta}{\dvol}}(A + A^T) = \sqrt{\frac{1}{2} \frac{\mu
\theta}{\dvol}}\mathcal{T}A$.   This gives $R_{visco} = \left(\sqrt{\frac{1}{2}
\frac{\mu \theta}{\dvol}}\right)\mathcal{T}$.  We can readily verify that  $R_{visco} R_{visco}^T =
\frac{1}{2} \frac{\mu \theta}{\dvol} \mathcal{T}^2 = \frac{\mu \theta}{\dvol}
\mathcal{T} = K_{visco}$.

For the $R_{heat}$, we have $K_{heat} = -\mbox{\small div}\left( \tilde{\kappa}
\nabla\right)$.  This yields factorization $R_{heat} = -\mbox{\small div}\left(
\sqrt{\tilde{\kappa}}\right)= (-\nabla \cdot)\sqrt{\tilde{\kappa}} I $, where
we use $(-\nabla \cdot)^T = \nabla$ and $R_{heat}R_{heat}^T = K_{heat}$.  We
see with these terms that $RR^T = K^{(2)}$ and we can generate the fluctuating
hydrodynamic contributions to the stochastic driving fields by using $\zeta =
\sqrt{2k_B} R\xi$.

For the fluid-particle interface, we split the matrix $K^{(3)}$ into two decoupled parts as 
\begin{eqnarray*}
K^{(3)} = K_1^{(3)} + K_2^{(3)}.
\end{eqnarray*}
The $K_1^{(3)}$ correspond to the dissipation from the mechanics and the 
$K_2^{(3)}$ to the direct heat exchanges.   We let 
\begin{eqnarray}
\label{equ_K_j_M_E_part1}
\\
\nonumber
K_1^{(3)} = 
\begin{bNiceMatrix}[last-row=5,last-col=5]
\theta_ID_I(Y) & 0 & 0 & \frac{-\theta_ID_I(Y)\dot{q}}{c_I} & \atzmlabel{$p_{P,F}$} \\
0 & 0 & 0 & 0& \atzmlabel{$\theta_P$}  \\
0 & 0 &  0 & 0& \atzmlabel{$\theta_F$}  \\ 
\frac{-\dot{q}_{P,F}^T\Theta(Y)D(Y)}{c_I} & 
0 &  0 & 
\frac{\dot{q}_{P,F}^T\Theta(Y)D(Y)\dot{q}_{P,F}}{c_{I,I}} & \atzmlabel{$\theta_I$} \\
\atzmlabel{$p_{P,F}$} & \atzmlabel{$\theta_P$} & \atzmlabel{$\theta_F$} & \atzmlabel{$\theta_I$} & 
\end{bNiceMatrix}.
\end{eqnarray} 
For consistency between the cases, we also show here some of the zero rows and columns. 
Similar to the particle case, we can decompose this into a factor of the form
\begin{eqnarray}
\label{equ_B_E__particle}
R_1 = 
\begin{bNiceMatrix}[last-row=5,last-col=2]
\sqrt{\theta_I}R_D(Y) & \atzmlabel{$p_{P,F}$} \\
0 & \atzmlabel{$\theta_P$} \\
0 & \atzmlabel{$\theta_F$} \\
-\frac{\sqrt{\theta_I}\dot{\mb{q}}_{P,F}^T R_D(Y)}{c_I } &
\atzmlabel{$\theta_I$} \\
p_{P,F} & \\ 
\end{bNiceMatrix}.
\end{eqnarray}
We need to derive $R_D$ so that $R_DR_D^T = D_I(Y)$.  This would yield
$R_1R_1^T= K_1^{(3)}$.  In the interface case, we have the 
dissipative tensor $D_I$ given by equation~\ref{equ_D_def}.  For this
purpose, we factor using
\begin{eqnarray}
R_D = 
\sqrt{\gamma}
\begin{bNiceMatrix}[last-row=3,last-col=2]
 I & \atzmlabel{$p_P$} \\ 
-\Lambda  & \atzmlabel{$p_F$} \\ 
 \atzmlabel{$p_P$} & \\ 
\end{bNiceMatrix}.
\end{eqnarray}
We remark that this only requires generating a random value that is the size of the 
particle degrees of freedom $p_F$.  This arises from the strong correlations required
to ensure the energy balance between the fluctuations that inter-convert between 
the fluid temperature fields and momentum fields.  \\

For the second factor corresponding to the heat exchanges, we have 
\begin{eqnarray}
\label{equ_K_j_M_E_part2}
\\
\nonumber
K_2^{(3)} = 
\begin{bNiceMatrix}[last-row=5,last-col=5]
0 & 0 & 0 & 0 & \atzmlabel{$p_{P,F}$} \\
0 & \frac{\kappa_{PI}\theta_I\theta_P}{c_{P,P}}& 0 & -\frac{\kappa_{PI}\theta_P\theta_I}{c_{P,I}}& \atzmlabel{$\theta_P$}   \\
0 & 0 &  \frac{\mbox{\tiny diag}(\kappa_{FI}\dvol\;\theta_F\theta_I)}{c_{F,F} \dvol \dvol} & -\frac{\kappa_{FI}\dvol\;\theta_I\theta_F}{c_{F,I} \dvol} & \atzmlabel{$\theta_F$}  \\ 
0 & 
-\frac{\kappa_{PI}\theta_I\theta_P}{c_{I,P}} & 
-\frac{(\kappa_{FI}\dvol\;\theta_I\theta_F)^T}{c_{I,F} \dvol}
&  
\frac{\kappa_{PI}\theta_P\theta_I + \theta_I \int \kappa_{FI} \theta_F dx}{c_{I,I}} & \atzmlabel{$\theta_I$} \\
\atzmlabel{$p_{P,F}$} & \atzmlabel{$\theta_P$} & \atzmlabel{$\theta_F$} & \atzmlabel{$\theta_I$} &  \\
\end{bNiceMatrix}.
\end{eqnarray}
We again retained some of the zero rows and columns for comparison with the
other parts of the decomposition.  For the continuum fluid fields, we split this
further into a decomposition of the form $K_{2}^{(3)} = K_{21}^{(3)} +
\int K_{22}^{(3)}(x) dx$.  In the discrete case, we would replace the integrals 
by the finite volume sums of the form $\sum_m K_{22}^{(3)}(x_m) \dvol$.  For 
clarity and to avoid clutter in expressions, 
we retain the integral notation.   We can factor 
these terms as
\begin{eqnarray}
\label{equ_K_j_M_E_part21}
\\
\nonumber
K_{21}^{(3)} &=& 
\begin{bNiceMatrix}[last-row=3,last-col=3]
\frac{\kappa_{PI}\theta_I\theta_P}{c_{P,P}}& -\frac{\kappa_{PI}\theta_P\theta_I}{c_{P,I}}& \atzmlabel{$\theta_P$}   \\
-\frac{\kappa_{PI}\theta_I\theta_P}{c_{I,P}} & 
\frac{\kappa_{PI}\theta_P\theta_I dx}{c_{I,I}} & \atzmlabel{$\theta_I$} \\
\atzmlabel{$\theta_P$} & \atzmlabel{$\theta_I$} & \atzmlabel{$(\mb{e}_P,\mb{e}_I)$} \\
\end{bNiceMatrix} \\
\nonumber
& = &  
\kappa_{PI}\theta_P\theta_I
\left[
\frac{1}{c_{P,P}} \mb{e}_P \mb{e}_P^T 
- 
\frac{1}{c_{P,I}} \mb{e}_P \mb{e}_I^T 
-
\frac{1}{c_{I,P}} \mb{e}_I \mb{e}_P^T 
+ 
\frac{1}{c_{I,I}} \mb{e}_I \mb{e}_I^T 
\right] \\
\nonumber
&=& R_{21}R_{21}^T.
\end{eqnarray}
The $\mb{e}_I$ and $\mb{e}_P$ give the standard basis vectors
with all zero entries except for the components corresponding
to the $\theta_I$ and $\theta_P$ degrees of freedom of the 
system.  This gives the factor 
\begin{eqnarray}
\label{equ_K_j_M_E_R21}
\\
\nonumber
R_{21} &=& 
\sqrt{\kappa_{PI}\theta_I\theta_P}
\left[
\begin{array}{r}
\frac{1}{c_{P}}\mb{e}_P\\
-\frac{1}{c_{I}}\mb{e}_I
\end{array}
\right]
= \sqrt{\kappa_{PI}\theta_I\theta_P}
\begin{bNiceMatrix}[last-row=3,last-col=2]
\frac{1}{c_{P}} & \atzmlabel{$\theta_P$} \\
-\frac{1}{c_{I}} & \atzmlabel{$\theta_I$} \\
& \\
\end{bNiceMatrix}.
\end{eqnarray}
We also expressed this factor using our notation for non-zero entries. 
As we will see in our derivations below, it is useful to have flexibility
between these ways of describing the system. 

The second factor is similar for each spatial location since 
$\kappa_{FI} = \kappa_{FI}(x)$ with
\begin{eqnarray}
\label{equ_K_j_M_E_part21}
\\
\nonumber
K_{22}^{(3)} &=&
\int 
\begin{bNiceMatrix}[last-row=3,last-col=3]
\frac{\kappa_{FI}\theta_I\theta_{F(x)}}{c_{F,F} \dvol \dvol}& -\frac{\kappa_{FI}\theta_{F(x)}\theta_I}{c_{F,I} \dvol} & \atzmlabel{$\theta_F(x)$} \\
-\frac{\kappa_{FI}\theta_I\theta_{F(x)}}{c_{I,F} \dvol} & 
\frac{\kappa_{FI}\theta_{F(x)}\theta_I }{c_{I,I}} & \atzmlabel{$\theta_I$} \\
\atzmlabel{$\theta_F(x)$} & \atzmlabel{$\theta_I$} & \atzmlabel{$(\mb{e}_{\theta_F(x)},\mb{e}_{\theta_I})$} \\
\end{bNiceMatrix}dx \\
\nonumber
&=& 
\int 
\kappa_{FI}\theta_{F(x)}\theta_I
\left[
\begin{array}{c}
\frac{1}{c_{F} \dvol} \mb{e}_{\theta_F(x)} \\
\nonumber
-\frac{1}{c_{I}} \mb{e}_{\theta_I}
\end{array} 
\right]
\left[
\begin{array}{c}
\frac{1}{c_{F} \dvol} \mb{e}_{\theta_F(x)} \\
-\frac{1}{c_{I}} \mb{e}_{\theta_I}
\end{array} 
\right]^T
dx \\
\nonumber
& = & \int R_{22}(x)R_{22}^T(x) dx,
\end{eqnarray}
where 
\begin{eqnarray}
\label{equ_K_j_M_E_R22}
R_{22}(x) &=& 
\sqrt{\kappa_{FI}(x)\theta_I\theta_F(x)\dvol}
\left[
\begin{array}{l}
\frac{1}{c_{F}\dvol} \mb{e}_{\theta_F(x)} \\
-\frac{1}{c_{I}} \mb{e}_{\theta_I}
\end{array} 
\right] \\
&=& 
\sqrt{\kappa_{FI}(x)\theta_I\theta_F(x)\dvol}
\begin{bNiceMatrix}[last-row=3,last-col=2]
\frac{1}{c_{F}\dvol} & \atzmlabel{$\theta_F(x)$}  \\
-\frac{1}{c_{I}} & \atzmlabel{$\theta_I$} \\
&  \\
\end{bNiceMatrix}.
\end{eqnarray}
The fluctuations are generated using
\begin{eqnarray}
\label{equ_M_E__R_0}
\\
\nonumber
\mb{g}_{2} = \mb{g}_{21} + \int \mb{g}_{22}(x)dx, \;\;
\mb{g}_{21} = \sqrt{k_B\Delta{t}}R_{21} \bsy{\xi}_{21},\;\;
\mb{g}_{22}(x) = \sqrt{k_B\Delta{t}}R_{22}(x) \bsy{\xi}_{22}(x).
\end{eqnarray}
As mentioned above, for brevity we express things in terms of integrals, but 
in practice, we replace these expressions by the finite volume sums 
$\int (\cdot) dx \sim \sum_m (\cdot)_{x_m} \dvol$.  This yields the needed correlations
in the fluctuations for $K^{(3)}$.  
Since the $\bsy{\xi}_{ij}(x_1),\bsy{\xi}_{ij}(x_2)$ have zero
correlation when $x_1 \neq x_2$.  We further have 
\begin{eqnarray}
\\
\nonumber
\langle 
\mb{g}_{2}
\mb{g}_{2}^T
\rangle
&=& \langle \mb{g}_{21} \mb{g}_{21}^T\rangle
+ 
\int 
\langle \mb{g}_{22} \mb{g}_{22}^T\rangle
dx \\
\nonumber
&=& 2 k_B R_{21}R_{21}^T\Delta{t} + \int 2 k_B R_{22}(x)R_{22}(x)^T dx \Delta{t}  \\
\nonumber
&=&
2 k_BK_{21}^{(j)} \Delta{t}  
 +  2 k_B\int K_{22}^{(j)}(x) dx \Delta{t} =
2 k_B K_{2}^{(3)}\Delta{t}.
\end{eqnarray}
This provides stochastic driving fields with the required statistics.  We can
combine these factors into one large $B$ matrix as follows.  If $\mb{g} =
\sqrt{2 k_B \Delta{t}} R_1 \bsy{\xi}_1 + \sqrt{2 k_B \Delta{t}} R_2
\bsy{\xi}_2$ then $\mb{g} = \sqrt{2k_B \Delta{t}}[R_1|R_2] [\bsy{\xi}_1 |
\bsy{\xi}_2]^T$.  We similarly use this with $\mb{g}_1 = \sqrt{2 k_B \Delta{t}}
R_1 \bsy{\xi}_1$ to obtain the 
thermal force $\mb{g} = \mb{g}_1 + \mb{g}_2$. 

We have now derived analytically factors $B^{(j)}$ 
for each $K^{(j)}$.   This provides our algorithms for
generating the stochastic driving fields $\mb{G}_{thm}(t)$ and
$\mb{g}_{thm}(x,t)$ in equation~\ref{equ_full_model}.
This allows for simulating thermal effects and fluctuations 
arising from the temperature gradients, hydrodynamics, 
and particle drift-diffusion taking into account the
correlations from the fluid-structure coupling and thermal exchanges
of heat. 

\subsection*{Spatial Discretizations for the Operators}
We briefly discuss how we obtain numerical approximations that preserve
needed structure in the operators.  For the gradient $\nabla$ 
and divergence $\nabla \cdot$, we use discretizations based on finite 
volume methods with 
$\mathcal{G}_{\mb{m}}^{(d)}\mb{u} 
= 
\left({\mb{u}_{\mb{m} + \mb{e}_d} - \mb{u}_{\mb{m} - \mb{e}_{d}}}\right)/{2\Delta{x}}.
$
The $\mb{m}=(m_1,m_2)$ gives the lattice index, $\Delta{x}$ the mesh spacing,
and $\mb{e}_d$ is the standard basis vector in direction $d$.  For the 
divergence we use $\mathcal{D}\cdot = -\mathcal{G}^T$. 
For the thermal exchanges in the fluid we discretize using directly 
Fourier's law for the finite volumes. In this initial 
work, we used for simplicity the most basic methods based on 
a uniform lattice and central differences.  Given some of the limitations 
of these types of discretizations, alternatives also can be developed using 
finite volume methods on staggered meshes, or other spatial discretizations
based on spectral or finite element methods
\cite{faires2012numerical,brenner2008mathematical,donev2010accuracy,delong2013temporal,
AtzbergerFluctHydroConfined2018,AtzbergerRD2010}.  
A key property we use in
our discretizations is to preserve the adjoint relations between the 
gradient and divergence operators.  To obtain 
discretizations for our stochastic numerical methods, we replace in our
derivations the terms $\nabla$ with $\mathcal{G}$ and 
$\nabla\cdot = \mbox{div}$ with $\mathcal{D}$.  We also replace 
the spatial integrals $\int (\cdot) dx$ by the finite volume sums 
$\sum_m (\cdot)_{x_m} \dvol$.  This provides 
spatial discretizations that preserve key properties of the operators.
This approach provides viable stochastic numerical methods 
for performing simulations of the particle-fluid system incorporating 
fluctuations.

\section*{Results}

\subsection*{Simulation Studies} 
As a demonstration of the approaches, we perform a few basic simulations showing some of the 
thermal effects and related phenomena that can be captured by the methods.  
We consider first a particle that has
been externally heated, such as excitation from exposure to a laser, which then cools by transferring 
heat to the interface.  We then consider an interface that has a spatially varying 
temperature gradient and simulate its impact on hydrodynamic fluctuations.

\subsubsection*{Particle Heating of the Interface}
\begin{figure}[h]

\centerline{\includegraphics[width=0.95\columnwidth]{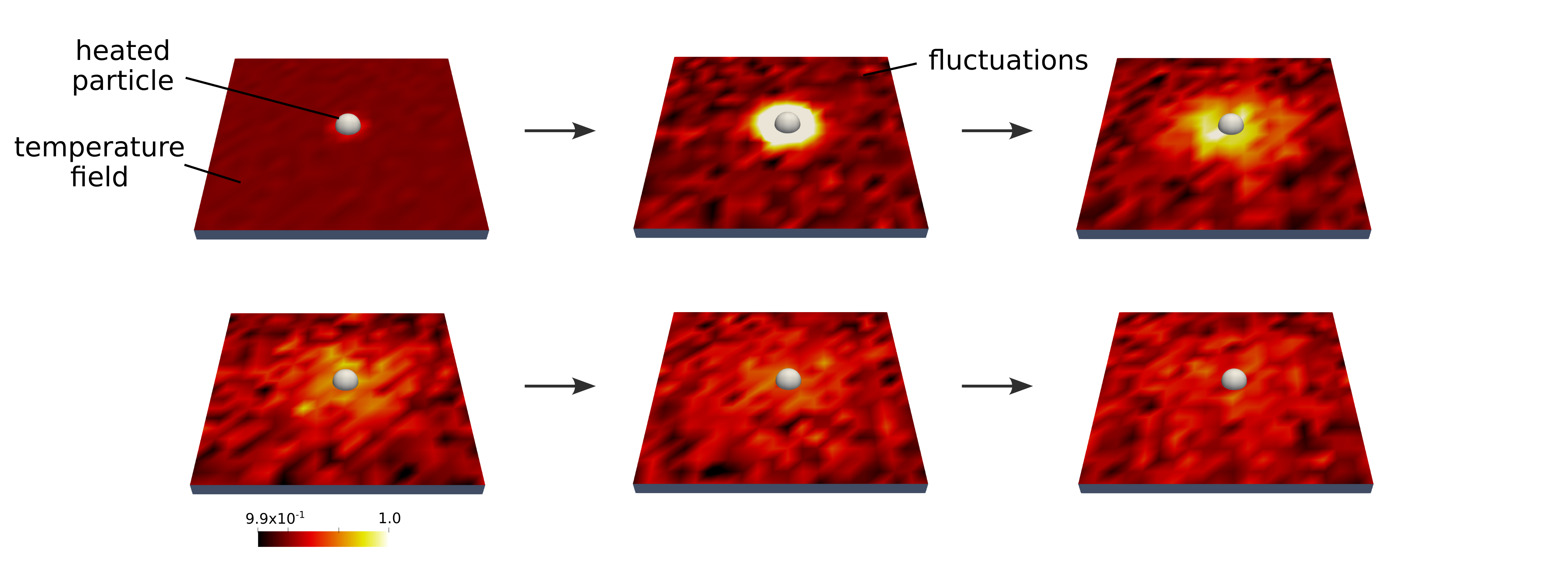}}
\caption{{\bf Particle Heating of the Interface.}
We show the simulation results for a particle that is externally driven to be
at an initially higher temperature and then cools to heat the surrounding
regions.  The heat transfer operators are based on $\Lambda$ operator which
give the thermal exchanges near the particle location $X$ through a spatially
varying thermal conductivity $\kappa_{F,I}(x;X)$.
The heat disperses according to Fourier's law 
with stochastic fluxes associated with spontaneous fluctuations.
\label{fig_particle_heating}}
\end{figure}

In recent experiments, particles within fluid interfaces are manipulated 
using lasers and other external driving fields inducing 
thermal effects~\cite{Jiang2010,Duhr2006,Chen2021,
jerabek2014microscale,piazza2008thermophoresis}.  As a basic model, we consider the 
energy exchanges of an initially heated particle that then cools by transferring 
heat to the surrounding regions of the interface.  This heat is further 
dispersed from Fourier's law laterally within the interface.  We show simulation
results for the temperature fields and fluctuations in 
Figure~\ref{fig_particle_heating}.

\begin{figure}[h]

\centerline{\includegraphics[width=0.95\columnwidth]{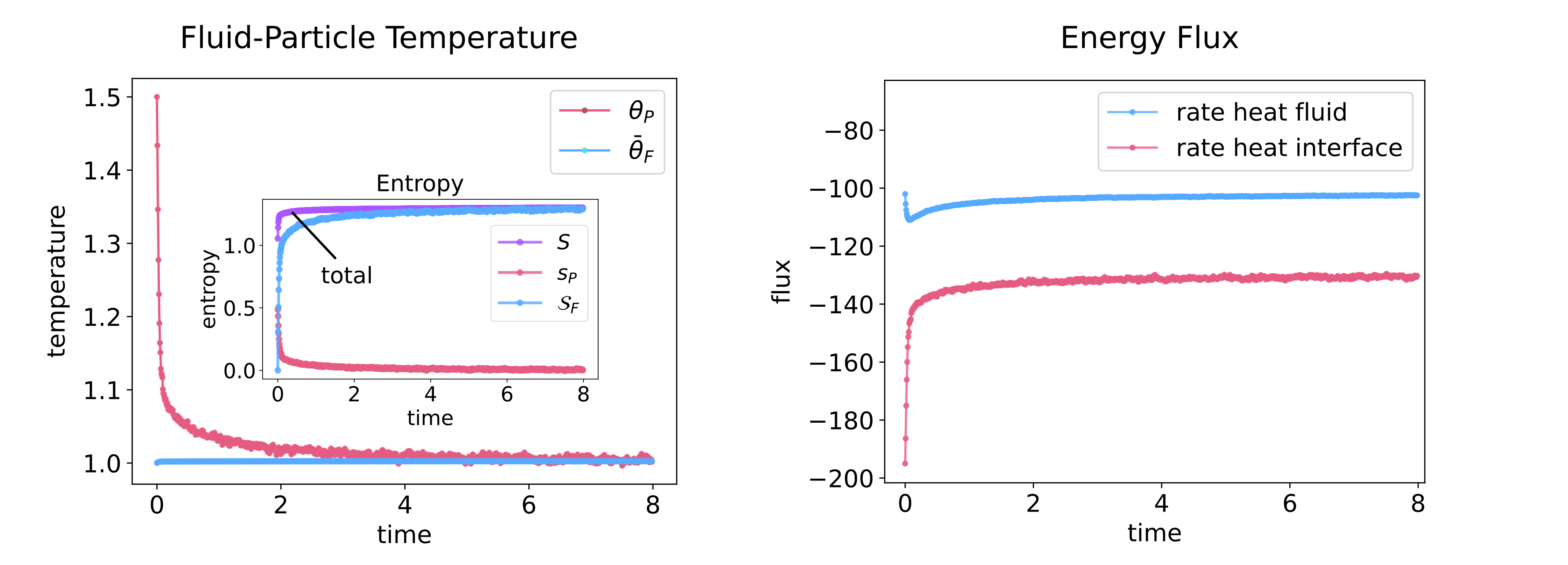}}
\caption{{\bf Particle Heating of the Interface.}   We show the results of the fluid-particle 
system where a hot particle heats the fluid interface.  The temperature of the particle 
and the average temperature of the fluid is shown on the \textit{(left)}.  We also show
the individual entropies of the particle, fluid, and the total entropy in the inset.
The energy exchanges between the particle and the fluid are shown on the \textit{(right)}. 
}
\label{fig_particle_heating_results}
\end{figure}

\begin{table}[ht]
\centering
{
\fontsize{9}{12}\selectfont
\begin{tabular}{|l|l|l|l|l|l|}
\hline 
\rowcolor{atz_table2}
\textbf{Parameter} & \textbf{Value} & \textbf{Parameter} & \textbf{Value} &
\textbf{Parameter} & \textbf{Value} \\
\hline
$\rho$ & $0.9$ & $\mu$ & $0.08$ & $m$ & $1.1$ \\
\hline
$\gamma$ & $5$ & $k_B$ & $10^{-5}$ & $\theta_P(0)$ & 1.5 \\
\hline
$\kappa_0$ & $4.2\times 10^6$ & $\kappa_{P,I}$ & $130$ & $\kappa_{F,I}$ & $102$ \\
\hline
$c_P$ & $1.2$ & $c_F$ & $130$ & $c_I$ & $1.4$ \\
\hline
$n_x$ & 20 & $n_y$ & 20 & $\Delta{x}$ & $10^{-1}$\\
\hline
$\Delta{t}$ & $10^{-3}$ & $n_t$ & $8000$ & $\dvol$ & $10^{-2}$ \\ 
\hline
\end{tabular}
} 
\caption{\textbf{Parameters.} For simulations of the case when the particle
heats the interface. The $n_x,n_y$ give the number of mesh sites in each
direction, $\Delta{x}$ the spatial discretization, $\Delta{t}$ the time-step,
$n_t$ the total number of time-steps.  The other parameters are discussed
in the context of the model equations~\ref{equ_full_model}.} 
\label{table_particle_heating}
\end{table}
When the particle is hottest there is initially a rapid transfer of energy from
the particle to the interface region.  Heat accumulates in the interfacial
region in the vicinity of the particle making the temperatures comparable.
This then decreases the rate of energy transfer from the particle 
with the heat transferred laterally within the interface.  Over time the particle and
interface equilibrate toward a common spatially uniform temperature.
As the fluid heats up this also results in increased 
fluctuations in the temperature fields.
We show quantitative results for the fluid-particle system for the temperature and 
energy exchanges over time in
Figure~\ref{fig_particle_heating_results}.

\clearpage
\subsubsection*{Temperature Gradients and Hydrodynamic Fluctuations}

\begin{figure}[h]

\centerline{\includegraphics[width=0.95\columnwidth]{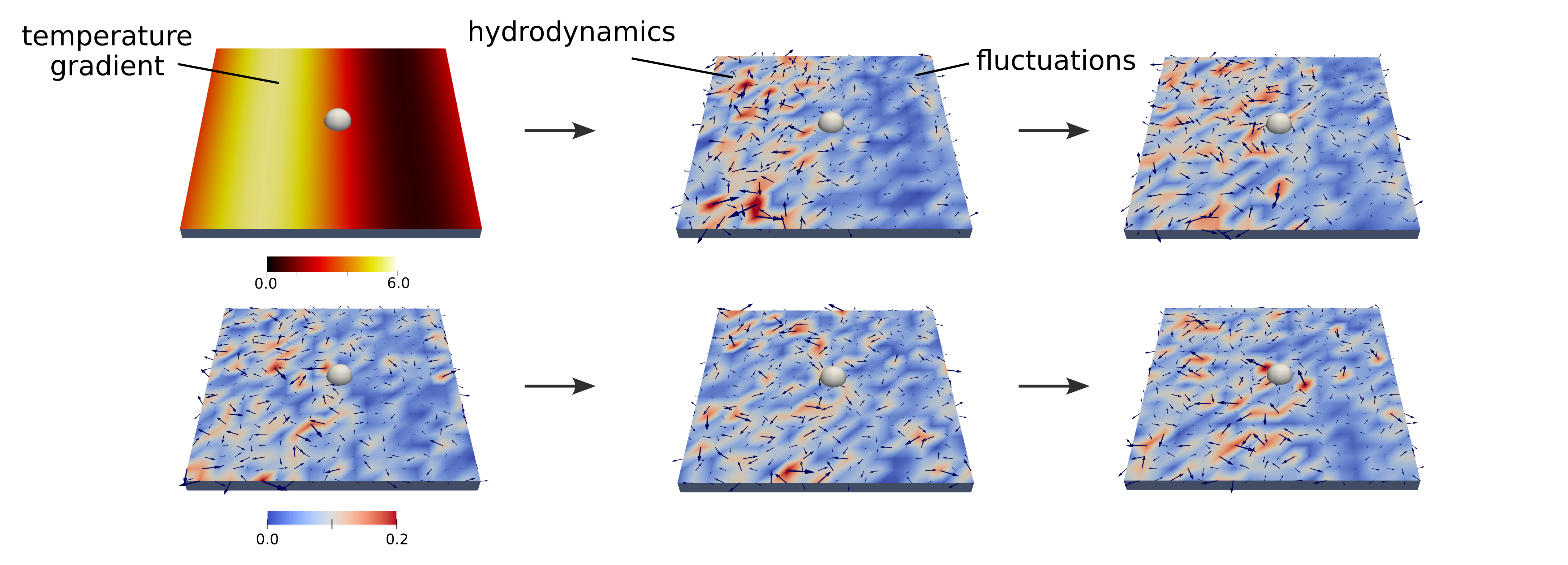}}
\caption{{\bf Hydrodynamic Fluctuations in a Temperature Gradient.}
The simulation results show the coupling where 
regions having larger temperature exhibit larger 
spontaneous hydrodynamic fluctuations.  These spatial variations in 
fluctuations drive inhomogeneous diffusivity of the particles and other 
statistical effects.  As the temperatures equilibrate and the gradient diminishes, the
spontaneous hydrodynamic fluctuations exhibit more spatial uniformity.
}
\label{fig_temp_grad_hydro}
\end{figure}

As another basic demonstration indicating some of the phenomena that
can be captured with the methods, we perform simulations when the
interface initially has a significant temperature gradient.  We show
these results in Figure~\ref{fig_temp_grad_hydro}.

\begin{table}[ht]
\centering
{
\fontsize{9}{12}\selectfont
\begin{tabular}{|l|l|l|l|l|l|}
\hline 
\rowcolor{atz_table2}
\textbf{Parameter} & \textbf{Value} & \textbf{Parameter} & \textbf{Value} &
\textbf{Parameter} & \textbf{Value} \\
\hline
$\rho$ & $0.9$ & $\mu$ & $0.08$ & $m$ & $1.1$ \\
\hline
$\gamma$ & $5$ & $k_B$ & $10^{-5}$ & $\theta_P(0)$ & 3.0 \\
\hline
$\kappa_0$ & $4.2\times 10^6$ & $\kappa_{P,I}$ & $130$ & $\kappa_{F,I}$ & $102$ \\
\hline
$c_P$ & $1.2$ & $c_F$ & $130$ & $c_I$ & $1.4$ \\
\hline
$n_x$ & 20 & $n_y$ & 20 & $\Delta{x}$ & $10^{-1}$\\
\hline
$\Delta{t}$ & $10^{-3}$ & $n_t$ & $16,000$ & $\dvol$ & $10^{-2}$ \\ 
\hline
\end{tabular}
} 
\caption{\textbf{Parameters.} For simulations of the case of the temperature gradient
and hydrodynamic fluctuations.
The $n_x,n_y$ give the number of mesh sites in each
direction, $\Delta{x}$ the spatial discretization, $\Delta{t}$ the time-step,
$n_t$ the total number of time-steps.  The other parameters are discussed
in the context of the model equations~\ref{equ_full_model}.} 
\label{table_temp_grad}
\end{table}

\begin{figure}[h]

\centerline{\includegraphics[width=0.95\columnwidth]{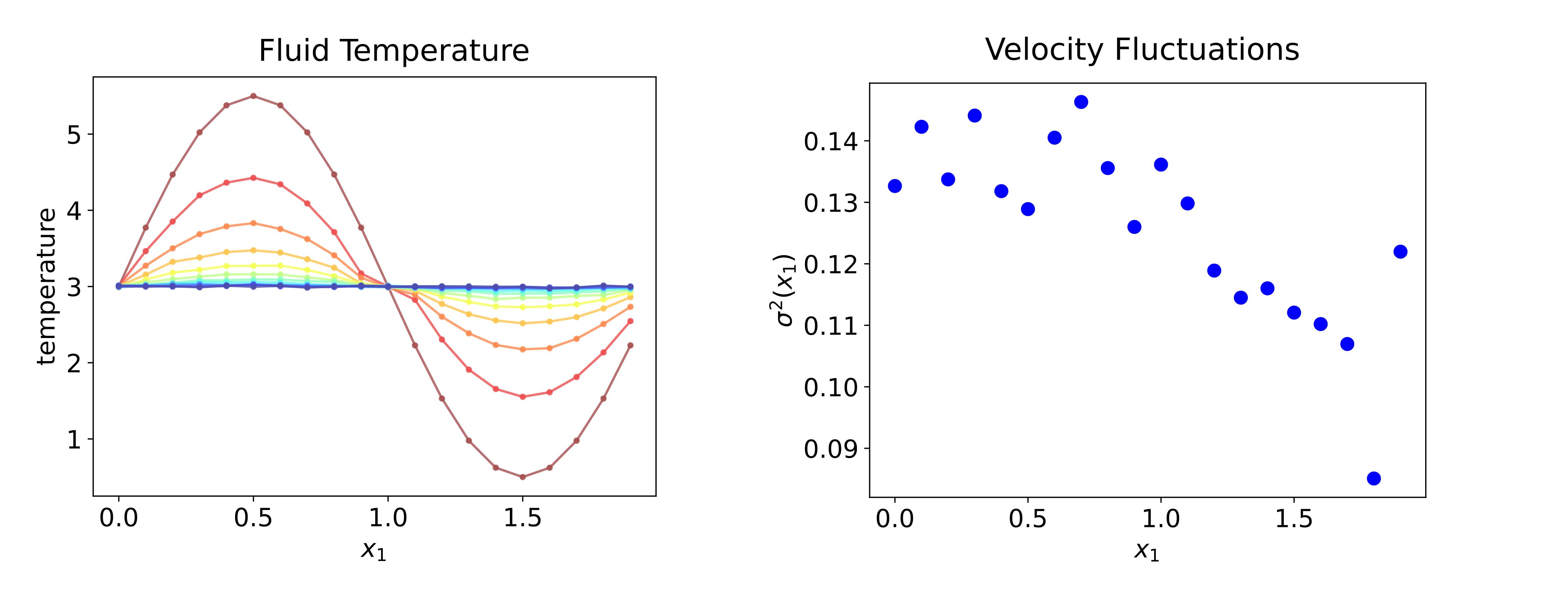}}
\caption{{\bf Hydrodynamic Fluctuations in a Temperature Gradient.}  
We show the temperature distribution as a function of $x_1$ and how it changes
over time on the \textit{(left)}.  From the lateral transfer of heat the 
temperature distribution becomes more uniform over time. 
We show estimates of the average variance $\sigma^2(x_1)$ of the
hydrodynamic fluctuations over the first $8,000$ simulation steps 
as a function of $x_1$ on the \textit{(right)}.  The variance of 
the fluctuations is seen to be smaller in the regions of lower temperature.
}
\label{fig_temp_grad_hydro_results}
\end{figure}

We see that the regions having the larger temperature exhibit hydrodynamic
fluctuations having larger variance as one may expect.  
This results in the temperature gradient
creating a hydrodynamic environment for the particles that results in larger
fluctuations and diffusivities in the larger temperature regions.  These
asymmetries can cause statistical effects in the drift-diffusion of
particles, which are related to the local hydrodynamic 
fluctuations~\cite{Atzberger2007,AtzbergerTabak2015}.  We show quantitative
results for the temperature fields and the variance $\sigma^2(\mb{x})$ of the hydrodynamic fluctuations in
Figure~\ref{fig_temp_grad_hydro_results}.

These simulation results demonstrate how the methods can be used simultaneously 
for thermal effects taking into account non-uniform temperature fields, 
fluctuations in the hydrodynamics, and the particle drift-diffusion dynamics. 
The methods can be used for further investigations in  active soft materials, 
complex fluids, and biophysical systems.

\clearpage

\section*{Conclusions}
We developed theory and modeling approaches for investigating the
non-equilibrium statistical mechanics of particle inclusions within fluid
interfaces.  Our approaches allow for taking into account the energy exchanges,
hydrodynamic coupling, and correlated spontaneous fluctuations of the
non-uniform temperature fields, fluid momentum fields, and the particle
drift-diffusion dynamics.  We developed practical numerical methods for
spatially discretizing the system and for efficiently generating the stochastic
driving fields yielding the correlated fluctuations.  We performed analysis to
obtain stochastic algorithms based on analytic factorizations of the operators.
In this initial work, we presented some basic demonstrations for how the
methods can be used to capture thermal effects for particles heating the
interface and for how hydrodynamic fluctuations are impacted by temperature
gradients.  The approaches provide methods for  modeling and simulation of
non-equilibrium statistical mechanics of thermal effects and fluctuations in
active and passive fluid-particle systems.  The simulation methods can be used
for further investigations in  active soft materials, complex fluids, and
biophysical systems.

\section*{Acknowledgments}
Authors PJA and DJ were supported by supported by NSF Grant DMS-1616353
and NSF Grant DMS-2306345. 
Author PJA also acknowledges UCSB Center for Scientific Computing NSF MRSEC
(DMR1121053) and UCSB MRL NSF CNS-1725797.  

\newpage
\clearpage

\printbibliography

\end{document}